# Near-Optimal Online Algorithms for Dynamic Resource Allocation Problems[*]

Patrick Jaillet[‡]        Xin Lu[§]

August 2012

## Abstract

In this paper, we study a general online linear programming problem whose formulation encompasses many practical dynamic resource allocation problems, including internet advertising display applications, revenue management, various routing, packing, and auction problems. We propose a model, which under mild assumptions, allows us to design near-optimal learning-based online algorithms that do not require the a priori knowledge about the total number of online requests to come, a first of its kind. We then consider two variants of the problem that relax the initial assumptions imposed on the proposed model.

## 1   Introduction

Online optimization is attracting wide attention from computer science and operations research communities. It has many applications, including those dealing with dynamic resource allocation problems. In many real-world problems, information about the instance to optimize is not completely known ahead of time, but revealed in an online fashion. For example, in typical revenue management problems, customers arrive sequentially offering a price for a subset of commodities, e.g. multi-leg flights. The seller must make irrevocable decisions to accept or reject customers at their arrivals, and try to maximize long-term overall revenue while respecting various resource constraints. Another example is the so-called AdWords problem, also known as the display ads problem. From keyword search queries arriving online, the problem is to sequentially allocate ad slots to budget-constrained bidders/advertisers. Similar problems appear in online routing problems, online packing problems, online auctions, and various internet advertising display applications.

In this paper, we consider a general online linear programming that covers many of the examples mentioned above. To be precise about the problem, we need to introduce some notations. Let $I$ be a set of $m$ resources; associated with each resource $i \in I$ is a capacity $b_i$. The set of resources and their capacities are known ahead of time. Let $J$ be a set of $n$ customers; each customer has a set of options $O_j$ and arrival time $t_j$. We assume that every customer has a bounded number of options, i.e. there exists a constant $q$ such that $|O_j| \leq q$ for all $j$. Each option $o \in O_j$ has a value $\pi_{jo}$ and requires $a_{ijo}$ units of resources $i$ for each $i \in I$, also written $\mathbf{a_{jo}}$ as a vector of dimension $m$. The set of options $O_j$ and associated $(\pi_{jo}, \mathbf{a_{jo}})$ are revealed at time $t_j$ when customer $j$ arrives. Upon arrival, the online algorithm must decide immediately and irrevocably whether or not to satisfy the customer, and, if yes, which option to choose. The goal is to find a solution that maximizes the overall revenue from customers while respecting resource constraints. More precisely, we consider the following linear program:

$$
\begin{array}{lll}
\max & \sum_j \sum_{o \in O_j} \pi_{jo} x_{jo} & \\
s.t. & \sum_{j,o} a_{ijo} x_{jo} \leq b_i, & \forall i \\
& \sum_{o \in O_j} x_{jo} \leq 1, & \forall j \\
& x_{jo} \geq 0, & \forall j, o
\end{array}
\tag{1}
$$

where $\forall j, \boldsymbol{\pi}_j \in (0,1]^{|O_j|}, \mathbf{a_j} \in [0,1]^{m \times |O_j|}$, and $\mathbf{b} \in \mathbb{R}_+^m$. In the online version of this problem, $(\boldsymbol{\pi}_j, \mathbf{a_j})$ is revealed only when customer $j$ arrives at time $t_j$. Upon that arrival, and constrained by irrevocable decisions $x_{j'o}$ made for customers arriving earlier, the online algorithm must then make decisions $x_{jo}$, such that

$$
\begin{array}{lll}
& \sum_{j': t_{j'} \leq t_j} \sum_{o \in O_{j'}} a_{ij'o} x_{j'o} \leq b_i, & \forall i \\
& \sum_{o \in O_j} x_{jo} \leq 1 & \\
& x_{jo} \geq 0, & \forall o \in O_j
\end{array}
\tag{2}
$$

---

[*]Research funded in part by NSF (grant 1029603) and ONR (grants N00014-09-1-0326 and N00014-12-1-0033).

[‡]Department of Electrical Engineering and Computer Science, Laboratory for Information and Decision Systems, Operations Research Center, MIT, Cambridge, MA, 02139; jaillet@mit.edu

[§]Operations Research Center, MIT, Cambridge, MA, 02139; luxin@mit.edu

The goal is to choose the variables $\mathbf{x}$ such that the objective function $\sum_j \sum_{o \in O_j} \pi_{jo} x_{jo}$ is maximized.

Several models (on how online instances are chosen) can be used to evaluate online algorithms, including the adversarial model, the i.i.d. model with or without knowledge of distributions, and the random permutation model. In the adversarial setting, no further assumption is made on the model. In that case, no online algorithm can achieve better than $O(1/n)$ fraction of the optimal offline solution [3]. However, as the adversarial setting is too conservative, it is natural to consider stochastic models. In the i.i.d. model with known distribution about future customers, positive results have been obtained for various problems. For many practical problems, such an assumption may be too strong, and the i.i.d. model without knowledge of the distribution would be more suitable. A weaker model, but easier to analyze, the random permutation model, has been considered more frequently. In that model, the order of customers is a uniform random permutation, and many near-optimal results have been obtained for it.

In this paper, we propose a new model, closer to the random permutation model, but removing a fundamental, yet practically questionable, assumption behind it. In all results using the random permutation model, the exact knowledge about the total number of customers to come is a key assumption, essential for ensuring near-optimality results. Without such information, no non-trivial result can be achieved. In many practical settings, including all the applications discussed above, this assumption is however far from being realistic. We consider instead a more realistic and natural setting, initially using the following two assumptions (the consequences of the relaxations of these two initial assumptions will also be considered in our paper):

**Assumption 1.** *Customers have i.i.d. random arrival times.*

The assumption is reasonable in many practical problems where customers' arrival rates are homogeneous throughout time. Ignoring the specific arrival times, the order of customers is essentially equivalent to the random permutation model. Later in the paper, we will relax the assumption and take heterogeneity of arrival rates into account.

**Assumption 2.** *The distribution governing random arrival times is known to the online algorithm.*

The assumption is necessary to estimate the total number of customers in case no past data is available. However, as discussed in Section 4, if a limited amount of past data is available, this assumption is not needed anymore.

For simplicity in the presentation of this paper, we make two additional technical assumptions, which can be removed without compromising the validity of our results, as we explain below.

**Assumption 3.** *The arrival time is modeled as a continuous random variable.*

No matter what the nature of the original random variable is, we can add an auxiliary random variable $t_j^a$, uniformly distributed between $[0, 1]$ for every customer $j$ upon his arrival. We define a total ordering on pairs $(t_j, t_j^a)$ based on lexical order. Note that the order of customers is preserved except for those who arrive exactly at the same time. The artificial ordering imposed on these customers does not help an online algorithm.

**Assumption 4.** *There are no degeneracies among all points $\{(\pi_{jo}, \mathbf{a_{jo}})\}_{j,o}$ and $(0, \mathbf{0})$, i.e. no $m + 2$ points share the same $m$-dimensional hyperplane.*

If this is not the case, we can introduce a random perturbation on $\pi_{jo}$: every $\pi_{jo}$ is multiplied by an i.i.d. random variable uniformly distributed between $[1, 1 + \epsilon]$. After the perturbation, there are no degeneracies almost surely. On the other hand, because the perturbation is small enough, the optimal value of the solution is affected by no more than a multiple factor of $1 + \epsilon$.

## 1.1 Our Techniques and Contributions

The online algorithms proposed in our paper share similar ideas with some other papers [6][2][7]: the algorithms first observe (without making any allocation) customers arriving early over a given period of time, and solve an offline LP problem over those customers. The corresponding optimal dual solution then works as a pricing mechanism for making online allocations on the following set of customers. The dual prices are updated from time to time to depict customers' preference more accurately as time moves along. We prove that such algorithms are $1 - \epsilon$ competitive under several different scenarios if resource capacities are large enough.

Our paper significantly improves previous results by removing the need to know a priori the number of customers $n$, a critical assumption in [6][2][7]. To the best of our knowledge, this is the first attempt to do so. As pointed out in several papers, knowing $n$ is so essential that no near-optimal online algorithms can be obtained even under a probabilistic version of that assumption. So a new model, with near-optimal online algorithm aspiration, would need to introduce alternative assumptions.

We believe our model with arrival time fits reality better: In practice, the setting that an online problem is more likely to face typically involves a known fixed period of time over which the customers are considered, rather than a known fixed number of customers to come. The question that a company usually asks is how to maximize revenue over a given period of time instead of how to maximize revenue over a given fixed number of future customers. The arrival time of a customer is also more natural and informative than his rank order. Furthermore, our model is more flexible as it allows, depending on specific applications, various extensions which can better fit real-world scenarios. For example, in airline revenue management problems, business customers and casual customers have different price-sensitivity and arrival time. The random permutation model cannot capture the heterogeneity among customers well. In contrast, our model can easily be extended to such scenarios, as demonstrated in Section 5.

We first consider problems where the distribution of arrival time is known in advance. Although similar in form, the previous approaches with fixed number of customers do not address our model well. One could first estimate the number of total arrivals in the early stage, and then use the estimation for the fixed-number algorithm as in [2]. However, the performance of this approach depends on the quality of the estimation. In order to keep the loss due to the estimation below $\epsilon$-fraction, the estimation error must be within $\epsilon$-fraction. According to concentration laws, it requires the total number of customers be at least $O(1/\epsilon^3)$. Noting that $b_i$'s are only required to be $O(1/\epsilon^2)$, this new requirement on the total number of customers is quite restrictive. On the other hand, our approach works for any number of customers, even if the number is smaller than $O(1/\epsilon^2)$.

We then consider two scenarios for which our initial assumptions are relaxed. In the first scenario, we do not assume the knowledge of the exact distribution, but some past observations instead. Instead of estimating the cumulative distribution function (CDF) on every point, we only make estimation on only a few critical points. This approach requires much less data points than the naive one. In the second scenario, we consider heterogeneous customers.

## 1.2 Literature Review

Inspired by applications such as advertisement display, online matching, and allocation problems have been recently studied extensively in the operations research and computer science communities. Three different models have been considered: adversarial model, i.i.d. model, and random permutation model.

In an adversarial model, no information is known to online algorithms about customers or requests. Karp et al. [12] consider a bipartite matching problem, present a best possible algorithm, RANKING, with a competitive ratio of $1 - 1/e$. Aggarwal et al. [1] propose a $1 - 1/e$-competitive algorithm for a vertex-weighted version of the same problem. Mehta et al. [15] and Buchbinder et al. [5] propose two different best possible algorithms for the Adwords problem.

In the random permutation model, the set of customers is still unknown to the online algorithm, but the order in which customers arrive is a uniformly random permutation. Goel and Mehta [9] prove that a greedy algorithm is $1 - 1/e$ competitive for the AdWords problem. Devanur and Hayes [6] present a near optimal online algorithm for the same problem under mild assumptions. More recently, Agrawal et al. [2] and Feldman et al. [7] propose near optimal algorithms, based on similar ideas, for general linear programming problems and packing problems. Mehdian and Yan [13] and Karande et al. [11] simultaneously prove RANKING algorithm is 0.696-competitive for the bipartite matching problem. Mirrokni et al. [16] propose an algorithm that works well for the AdWords problem in both adversarial and random permutation model.

In the i.i.d. model, customers or requests are drawn repeatedly and independently from a known probability distribution. Feldman et al. [8] present a 0.670-competitive algorithm for a bipartite matching problem. Manshadi et al. [14] give a 0.702-competitive algorithm for a slight variation of the same problem. Jaillet and Lu [10] improve both these bounds to 0.729 and 0.706, respectively.

**Organization:** The rest of the paper is organized as follows. In Sections 2 and 3, we present online algorithms for our basic model and prove that they are near optimal under mild conditions. In the following two sections, we consider situations where we can remove assumptions imposed on the model: the assumption on the knowledge of arrival distributions in Section 4 and the assumption on the homogeneity of customers in Section 5. In both sections, we propose and prove near optimal online algorithms.

## 2 One-Time Learning

Let $F(\cdot)$ be the cumulative distribution function of the random arrival time of customers. Assumption 3 ensures that its inverse $F^{-1}(\cdot)$ is well defined. Consider $S_\epsilon = \{j : t_j \leq F^{-1}(\epsilon)\}$, the set of customers arriving earlier than $F^{-1}(\epsilon)$. From Assumption 1, every customer belongs to $S_\epsilon$ with probability $\epsilon$.

The online algorithm observes customers in $S_\epsilon$, rejects them all, and then computes dual prices by solving the following primal dual LPs:

$$
\begin{aligned}
&\max && \sum_{j \in S_\epsilon, o \in O_j} \pi_{jo} x_{jo} && \min && \sum_i (1-\epsilon) \epsilon b_i p_i + \sum_{j \in S_\epsilon} q_j \\
&s.t. && \sum_{j \in S_\epsilon, o \in O_j} a_{ijo} x_{jo} \le (1-\epsilon)\epsilon b_i, \forall i && s.t. && \sum_i a_{ijo} p_i + q_j \ge \pi_{jo}, \forall j \in S_\epsilon, o \in O_j \\
& && \sum_{o \in O_j} x_{jo} \le 1, \forall j \in S_\epsilon && && p_i \ge 0, \forall i \\
& && x_{jo} \ge 0, \forall j \in S_\epsilon, o \in O_j && && q_j \ge 0, \forall j \in S_\epsilon
\end{aligned}
\tag{3}
$$

Let $\hat{\mathbf{x}}$ and $\hat{\mathbf{p}}$ be the optimal primal and dual solutions for these problems.

For customers arriving later, they are accepted if payments exceed the threshold set by $\hat{\mathbf{p}}$:

$$
x_{jo}(\hat{\mathbf{p}}) = \begin{cases} 1, \text{ if } \pi_{jo} - \sum_i a_{ijo}\hat{p}_i > \max_{o' \neq o}\{\pi_{jo'} - \sum_i a_{ijo'}\hat{p}_i, 0\} \\ \\ 0, \text{ otherwise} \end{cases}
\tag{4}
$$

The proposed online algorithm is then as follows:

---
**Algorithm 1** Online Learning Algorithm(OLA)
---
1: Reject all customers arriving earlier than $F^{-1}(\epsilon)$.
2: Let $\mathbf{x_j} = \mathbf{x_j}(\hat{\mathbf{p}})$ for customers arriving after $F^{-1}(\epsilon)$.

---

We call a customer $j$ degenerate if the maximizer for $\max_o\{\pi_{jo} - \sum_i a_{ijo}\hat{p}_i\}$ is not unique or $\pi_{jo} - \sum_i a_{ijo}\hat{p}_i = 0$. Degeneracies may lead to undesired results. Fortunately, due to Assumption 4, there are at most $m + 1$ degenerate customers, and all of them, if any, are in $S_\epsilon$. For degenerate $j$, the decision rule $\mathbf{x_j}(\hat{\mathbf{p}}) = \mathbf{0}$. For non-degenerate $j$, using complementary slackness, $\mathbf{x_j}(\hat{\mathbf{p}})$ equals the optimal solution $\hat{\mathbf{x_j}}$ to LP (3), as stated in the following lemma (see proof in appendix):

**Lemma 1.** *For non-degenerate $j \in S_\epsilon$, $\hat{x}_{jo} = x_{jo}(\hat{\mathbf{p}})$ for all $o \in O_j$.*

In this paper, we repeatedly use concentration laws to show that some undesired events rarely happen. In particular, we use Bernstein inequalities:

**Bernstein Inequalities [4]:** Let $X_1, ..., X_n$ be independent zero-mean random variables. Suppose there exists $M > 0$ such that $|X_i| \le M$ almost surely for all $i$. Then, $\forall t$,

$$
\Pr(\sum_{i=1}^n X_i > t) \le \exp\Big(-\frac{t^2/2}{\sum_i \mathbb{E}[X_i^2] + Mt/3}\Big).
$$

Toward the analysis of our online algorithms, we first show that the resulting solution is feasible with high probability:

**Lemma 2.** *If $\min_i b_i \ge 5m \ln(nq/\epsilon)/\epsilon^3$, then w.p. $1 - \epsilon$, $\sum_{j,o} a_{ijo} x_{jo}(\hat{\mathbf{p}}) \le b_i$ for all $i$.*

*Proof.* From Lemma 1, we have

$$
\sum_{j \in S_\epsilon, o} a_{ijo} x_{jo}(\hat{\mathbf{p}}) \le \sum_{j \in S_\epsilon, o} a_{ijo} \hat{x}_{jo} \le \epsilon(1-\epsilon)b_i.
$$

We would like to apply Bernstein inequalities to show that $\sum_{j,o} a_{ijo} x_{jo}(\hat{\mathbf{p}}) \geq b_i$ rarely happens. The difficulty is that the random variables $\{a_{ijo} x_{jo}(\hat{\mathbf{p}})\}_{j,o}$ depend on the realization of $S$ via $\hat{\mathbf{p}}$. To get around the issue, let us first fix $\mathbf{p}$ and $i$, and consider the event

$$\{\sum_{j,o} a_{ijo} x_{jo}(\mathbf{p}) \geq b_i, \sum_{j \in S_\epsilon, o} a_{ijo} x_{jo}(\mathbf{p}) \leq (1-\epsilon)\epsilon b_i\} \tag{5}$$

For every customer $j$, because the arrival time is uniformly distributed between $[0, T]$, $j \in S_\epsilon$ with probability $\epsilon$. Hence,

$$\mathbb{E}[\sum_{j \in S_\epsilon, o} a_{ijo} x_{jo}(\mathbf{p})] = \epsilon \cdot \mathbb{E}[\sum_{j,o} a_{ijo} x_{jo}(\mathbf{p})]$$

Using Bernstein's inequalities, we have

$$\begin{aligned}
&\Pr(\sum_{j,o} a_{ijo} x_{jo}(\mathbf{p}) \geq b_i, \sum_{j \in S_\epsilon, o} a_{ijo} x_{jo}(\mathbf{p}) \leq (1-\epsilon)\epsilon b_i) \\
\leq\ &\Pr(\sum_{j,o} a_{ijo} x_{jo}(\mathbf{p}) \geq b_i, \sum_{j \in S_\epsilon, o} a_{ijo} x_{jo}(\mathbf{p}) - \mathbb{E}[\sum_{j \in S_\epsilon, o} a_{ijo} x_{jo}(\mathbf{p})] \leq -\epsilon^2 b_i) \\
\leq\ &\exp(-\epsilon^3 b_i/4) \leq \epsilon/mn^m.
\end{aligned}$$

According to [17], $\mathbb{R}^{|I|}$ can be divided into no more than $(nq)^m$ regions such that all $\mathbf{p}$ in a region lead to the same $\mathbf{x}(\mathbf{p})$. By taking union bounds over all possible $\mathbf{p}$ and $i$, we have with probability $\epsilon$, there exist $i$ and $\mathbf{p}$ such that (5) is true. So:

$$\begin{aligned}
&\Pr(\exists i, \sum_{j,o} a_{ijo} x_{jo}(\hat{\mathbf{p}}) \geq b_i) \\
=\ &\Pr(\exists i, \sum_{j,o} a_{ijo} x_{jo}(\hat{\mathbf{p}}) \geq b_i, \sum_{j \in S_\epsilon, o} a_{ijo} x_{jo}(\hat{\mathbf{p}}) \leq (1-\epsilon)\epsilon b_i) \\
\leq\ &\Pr(\exists i, \mathbf{p}, \sum_{j,o} a_{ijo} x_{jo}(\mathbf{p}) \geq b_i, \sum_{j \in S_\epsilon, o} a_{ijo} x_{jo}(\mathbf{p}) \leq (1-\epsilon)\epsilon b_i) \leq \epsilon.
\end{aligned}$$

By taking the complement, we conclude the lemma. □

After obtaining feasibility, we now compare the online solution with the offline optimal solution $OPT$. Note that $OPT$ is the optimal solution to LPs:

$$\begin{aligned}
\max\ &\sum_{j,o} \pi_{jo} x_{jo} & \min\ &\sum_i b_i p_i + \sum_j q_j \\
s.t.\ &\sum_{j,o} a_{ijo} x_{jo} \leq b_i, \forall i & s.t.\ &\sum_i a_{ijo} p_i + q_j \geq \pi_{jo}, \forall j, o \\
&\sum_{o \in O_j} x_{jo} \leq 1, \forall j & &p_i \geq 0, \forall i \\
&x_{jo} \geq 0, \forall j, o & &q_j \geq 0, \forall j
\end{aligned} \tag{6}$$

We now show that the objective value of online solution $\mathbf{x}(\hat{\mathbf{p}})$ is close to $OPT$:

**Lemma 3.** *If* $\min_i b_i \geq 5m \ln(nq/\epsilon)/\epsilon^3$, *then w.p.* $1 - \epsilon$,

$$\sum_{j,o} \pi_{jo} x_{jo}(\hat{\mathbf{p}}) \geq (1 - 3\epsilon)OPT \tag{7}$$

*Proof.* Let us consider the following LP:

$$\begin{aligned}
\max\ &\sum_{j,o} \pi_{jo} x_{jo} \\
s.t.\ &\sum_{j,o} a_{ijo} x_{jo} \leq \hat{b}_i, &&\forall i \\
&\sum_{o \in O_j} x_{jo} \leq 1, &&\forall j \\
&x_{jo} \geq 0, &&\forall j, o
\end{aligned} \tag{8}$$

where $\hat{b}_i = \sum_{j,o} a_{ijo} x_{jo}(\hat{\mathbf{p}})$, if $\hat{p}_i > 0$ and $\hat{b}_i = \max\{\sum_{j,o} a_{ijo} x_{jo}(\hat{\mathbf{p}}), b_i\}$, if $\hat{p}_i = 0$. By complementary slackness, $\mathbf{x}(\hat{\mathbf{p}})$ is the optimal solution to this LP.

We then show that with probability $1 - \epsilon$, $\hat{b}_i \geq (1 - 3\epsilon) b_i, \forall i$. For $i$ such that $\hat{p}_i = 0$, it is trivially true from the definition of $\hat{b}_i$. For $i$ such that $\hat{p}_i > 0$, by complementary slackness, we have $\sum_{j \in S_\epsilon, o} \pi_{jo} \hat{x}_{jo} = (1 - \epsilon)\epsilon b_i$. Furthermore, according to Assumption 4 and Lemma 1, at most $m + 1$ different $j$ make $x_{jo}(\hat{\mathbf{p}}) \neq \hat{x}_{jo}$. Noting that $\pi_{jo} \in (0, 1]$, we have

$$\sum_{j \in S_\epsilon, o} \pi_{jo} x_{jo}(\hat{\mathbf{p}}) \geq \sum_{j \in S_\epsilon, o} \pi_{jo} \hat{x}_{jo} - (m + 1) = (1 - \epsilon)\epsilon b_i - (m + 1) \geq (1 - 2\epsilon)\epsilon b_i$$

Using the same technique as in the proof of Lemma 2, we can show that

$$\Pr\left(\exists i, s.t. \sum_{j \in S_\epsilon, o} \pi_{jo} x_{jo}(\hat{\mathbf{p}}) \geq (1 - 2\epsilon)\epsilon b_i, \sum_{j \in S, o} \pi_{jo} x_{jo}(\hat{\mathbf{p}}) \leq (1 - 3\epsilon) b_i\right) \leq \epsilon.$$

Hence, with probability $1 - \epsilon$, $\hat{b}_i \geq (1 - 3\epsilon) b_i, \forall i$.

In that case, we argue that $\sum_{j,o} \pi_{jo} x_{jo}(\hat{\mathbf{p}}) \geq (1 - 3\epsilon) OPT$. In fact, assuming $\mathbf{x}^*$ is the optimal solution to LP (6), then $(1 - 3\epsilon)\mathbf{x}^*$ is feasible to LP (8). As the optimal solution to LP (8), $\hat{\mathbf{x}}(\hat{\mathbf{p}})$ is no worse than $(1 - 3\epsilon)\mathbf{x}^*$, which concludes the lemma. □

Note that the left hand side of (7) includes revenue from customers in $S_\epsilon$, which should be excluded. It is upper bounded by $OPT_\epsilon$, the optimal solution to LP (3).

**Lemma 4.** $\mathbb{E}[OPT_\epsilon] \leq \epsilon \cdot OPT$.

*Proof.* Note that the optimal dual solution $(\mathbf{p}^*, \mathbf{q}^*)$ to (6) is also feasible to the partial dual problem (3). Hence, the optimal solution to (3): $OPT_\epsilon \leq (1 - \epsilon)\epsilon \sum_i b_i p_i^* + \sum_{j \in S} q_j^*$. By taking expectation on both sides, we can conclude our lemma. □

We are now ready to prove the main theorem.

**Theorem 1.** *If* $\min_i b_i \geq 5m \ln(nq/\epsilon)/\epsilon^3$, *OLA is* $1 - O(\epsilon)$ *competitive.*

*Proof.* From the lemmas above, with probability $1 - 2\epsilon$ (denoted by $\mathcal{E}_1$), the solution $\mathbf{x}(\hat{\mathbf{p}})$ is feasible and $\sum_{j,o} \pi_{jo} x_{jo}(\hat{\mathbf{p}}) \geq (1 - 3\epsilon) OPT$. Then,

$$
\begin{aligned}
\mathbb{E}[\sum_{j \notin S_\epsilon, o \in O_j} \pi_{jo} x_{jo}(\hat{\mathbf{p}})] &\geq \mathbb{E}[\sum_{j \notin S_\epsilon, o \in O_j} \pi_{jo} x_{jo}(\hat{\mathbf{p}})|\mathcal{E}_1] \cdot \Pr(\mathcal{E}_1) \\
&\geq (\mathbb{E}[\sum_{j,o} \pi_{jo} x_{jo}(\hat{\mathbf{p}})|\mathcal{E}_1] - \mathbb{E}[OPT_\epsilon|\mathcal{E}_1]) \Pr(\mathcal{E}_1) \\
&\geq (1 - 3\epsilon)(1 - 2\epsilon) OPT - \epsilon OPT \geq (1 - 6\epsilon) OPT
\end{aligned}
$$

□

## 3 Dynamic Pricing Algorithm

The basic idea of OLA is to compute dual prices for resources, based on customers who arrive early. However, because of the limited number of customers, $\min_i b_i$ is required to be as large as $O(1/\epsilon^3)$ to have a small error probability. A natural question is if sampling more customers can help. The answer is affirmative as showed in this section.

Let $\epsilon = 2^{-E}$, where $E \in \mathbb{N}$. Let $S_l = \{j : t_j < F^{-1}(l)\}$ be the set of customers arriving no later than $F^{-1}(l)(l \in L = \{\epsilon, 2\epsilon, 4\epsilon, ...\})$. Let $\hat{\mathbf{p}}_\mathbf{l}$ denote the optimal dual solution to the following partial LPs:

$$
\begin{array}{lll}
\max & \sum_{j \in S_l, o \in O_j} \pi_{jo} x_{jo} & \\
s.t. & \sum_{j \in S_l, o \in O_j} a_{ijo} x_{jo} \leq (1 - h_l) lb_i, & \forall i \\
& \sum_{o \in O_j} x_{jo} \leq 1, & \forall j \in S_l \\
& x_{jo} \geq 0, & \forall j \in S_l, o \in O_j
\end{array}
\tag{9}
$$

where $h_l = \epsilon \sqrt{1/l}$.

Unlike OLA, DPA updates dual prices multiple times to have better performance:

---
**Algorithm 2** Dynamic Pricing Algorithm(DPA)
---
1: Reject all customers arriving earlier than $\epsilon T$.
2: Update dual prices $\hat{\mathbf{p}}_\mathbf{l}$ at time $\epsilon T, 2\epsilon T, 4\epsilon T, ...$
3: Let $x_j = x_j(\hat{\mathbf{p}}_\mathbf{l})$ for customers arriving between $lT$ and $2lT$.
---

The analysis of DPA is very similar to the one of OLA. We show that with high probability, the resulting solution is feasible, the resulting solution is near optimal, and the loss caused by observation process is small. Because of the lack of space, proofs are omitted here and can be found in the appendix.

**Lemma 5.** *If* $\min_i b_i \geq 10m \ln(nq/\epsilon)/\epsilon^2$, *then w.p.* $1 - \epsilon, \sum_{j \in S_{2l} \setminus S_l, o \in O_j} a_{ijo} x_{jo}(\hat{\mathbf{p}}_\mathbf{l}) \leq lb_i, \forall i, l$

**Lemma 6.** *If* $\min_i b_i \geq 10m \ln(nq/\epsilon)/\epsilon^2$, *then w.p.* $1 - \epsilon, \sum_{j \in S_{2l}, o \in O_j} \pi_{jo} x_{jo}(\hat{\mathbf{p}}_\mathbf{l}) \geq (1 - 2h_l - \epsilon)OPT_{2l}, \forall l$.

**Lemma 7.** *Let* $OPT_l$ *be the optimal value to (9), then* $\mathbb{E}[OPT_l] \leq l \cdot OPT$.

Combining Lemma 5, 6, and 7, we conclude the main result:

**Theorem 2.** *If* $\min_i b_i \geq 10m \ln(nq/\epsilon)/\epsilon^2$, *then DPA is* $1 - O(\epsilon)$ *competitive.*

# 4 Learning From the Past

The previous two sections discuss problems where the distribution of customers' arrival time is known to the online algorithm ahead of time. However, the assumption may not be true in many applications. Instead, the online algorithm is more likely to have access to past data rather than the exact distribution. For example, from observation on previous days, a retail store owner may expect that roughly two-thirds of the customers arrive in the afternoon. Specifically, in this section, we assume customers have i.i.d. arrival time with unknown distribution. Furthermore, information about the $k$ past customers $\{t'_k, \mathbf{a}'_\mathbf{k}, \boldsymbol{\pi}'_k\}$ is given to the online algorithm. The algorithm proposed in this section only uses arrival times of past customers.

Intuitively, by concentration laws, the distribution $f(\cdot)$ can be estimated arbitrarily well point-wise as $k$ grows. However, point-wise accuracy is unnecessary for our algorithm, and requires a huge amount of data. Note that, only at time $F^{-1}(\epsilon), F^{-1}(2\epsilon), F^{-1}(4\epsilon), ...$ does DPA update its pricing policy. Thus, if we could estimate those quantile points well, we would expect the resulting algorithm has similar performance as DPA.

First, let us show that $t'_{lk}$ is a good estimate of $F^{-1}(l)$. To be more precise,

**Lemma 8.** *If $k \geq 5\ln(1/\epsilon)/\epsilon^2$, w.p. $1 - \epsilon$, $F^{-1}((1 - h_l)l) \leq t'_{lk} \leq F^{-1}((1 + h_l)l), \forall l \in L = \{\epsilon, 2\epsilon, 4\epsilon, ...\}$. Here $h_l = \epsilon\sqrt{1/l}$.*

*Proof.* Let $N_l^1$ be the number of customers arriving between $[0, F^{-1}((1 - h_l)l)]$. Then,

$$\Pr(N_l^1 \geq lk) = \Pr(N_l^1 - \mathbb{E}[N_l^1] \geq h_l lk) \leq \exp(-\epsilon^2 k/2).$$

Let $N_l^2$ be the number of customers arriving between $[0, F^{-1}((1 + h_l)l)]$. Then,

$$\Pr(N_l^2 \leq lk) = \Pr(N_l^2 - \mathbb{E}[N_l^1] \leq -h_l lk) \leq \exp(-\epsilon^2 k/4).$$

Noting that $N_l^1 \geq lk$ is equivalent to $t'_{lk} \geq F^{-1}((1 - h_l)l)$ and $N_l^2 \leq lk$ is equivalent to $t'_{lk} \leq F^{-1}((1 + h_l)l)$. Therefore, by union bound, $F^{-1}((1 - h_l)l) \leq t'_{lk} \leq F^{-1}((1 + h_l)l), \forall l = \epsilon, 2\epsilon, 4\epsilon, ...$ w.p. $1 - 2\ln(1/\epsilon)\exp(-\epsilon^2 k/4) \geq 1 - \epsilon$. $\qquad\square$

After obtaining estimates of $F^{-1}(l)$, let us present the online algorithm DPAD. The only difference from DPA is that instead of updating at $F^{-1}(l)$, DPAD updates its pricing policy at $t'_{lk}$.

---

**Algorithm 3** Dynamic Pricing Algorithm with Data(DPAD)

1: Reject all customers arriving earlier than $t'_{\epsilon k}$.
2: Update dual prices $\hat{\mathbf{p}}_\mathbf{l}$ at time $t'_{\epsilon k}, t'_{2\epsilon k}, t'_{4\epsilon k}, ...$ according to LP (10) given below
3: Let $x_j = x_j(\hat{\mathbf{p}}_\mathbf{l})$ for customers arriving between $t'_{lk}$ and $t'_{2lk}$.

---

$$
\begin{aligned}
\max \quad & \sum_{j \in S_l, o \in O_j} \pi_{jo} x_{jo} \\
s.t. \quad & \sum_{j \in S_l, o \in O_j} a_{ijo} x_{jo} \leq (1 - 6h_l)lb_i, \quad \forall i \\
& \sum_{o \in O_j} x_{jo} \leq 1, \quad\quad\quad\quad\quad\quad\quad \forall j \in S_l \\
& x_{jo} \geq 0, \quad\quad\quad\quad\quad\quad\quad\quad\quad \forall j \in S_l, o \in O_j
\end{aligned}
\tag{10}
$$

where $h_l = \epsilon\sqrt{1/l}$ and $S_l = \{j : t_j \leq t'_{lk}\}$.

Let event $\mathcal{E}_{est}$ denote the event where $F^{-1}(\epsilon), F^{-1}(2\epsilon), F^{-1}(4\epsilon), ...$ are well-estimated as in Lemma 8. Given $\mathcal{E}_{est}$, we would expect DPAD has many similar properties as DPA. Indeed, it is the case, and the analysis is almost identical. We show that with high probability, the resulting solution is feasible, the resulting solution is near optimal, and the loss due to observation is small.

**Lemma 9.** *Given $\mathcal{E}_{est}$, if $\min_i b_i \geq 3m\ln(nq/\epsilon)/\epsilon^2$, then with probability $1 - \epsilon$,*

$$\sum_{j \in S_{2l}\setminus S_l, o \in O_j} a_{ijo} x_{jo}(\hat{\mathbf{p}}_\mathbf{l}) \leq lb_i, \forall i, l.$$

**Lemma 10.** *Given $\mathcal{E}_{est}$, if $\min_i b_i \geq 3m\ln(nq/\epsilon)/\epsilon^2$, then with probability $1 - \epsilon$,*

$$\sum_{j \in S_{2l}, o \in O_j} \pi_{jo} x_{jo}(\hat{\mathbf{p}}_\mathbf{l}) \geq (1 - 9h_l - \epsilon)OPT_{2l}, \forall l.$$

**Lemma 11.** *Given $\mathcal{E}_{est}$, $\mathbb{E}[OPT_l] \leq (1 + h_l)l \cdot OPT$.*

From lemmas above, we can conclude:

**Theorem 3.** *If $\min_i b_i \geq 3m\ln(nq/\epsilon)/\epsilon^2$ and $k \geq 5\ln\epsilon/\epsilon^2$, the algorithm is $1 - O(\epsilon)$ competitive.*

The assumptions made in the theorem are reasonable. On the one hand, the lower bound on $\min_i b_i$ is the same as in DPA, which has been showed to be best possible in many occasions. On the other hand, the lower bound on $k$ is even lower than the one on $\min_i b_i$, which means only a limited amount of past observations are required to obtained the near-optimal result.

DPAD does not take advantage of demands and prices information from past customers. In the random permutation model, such information is unlikely to improve the online algorithm. But for practical problems where demands and payments come from unknown i.i.d. distributions, these data may provide good estimation on the distributions, and lead to better results.

## 5   Heterogeneous Customers

As we may see in many applications, customers are not all homogeneous. Customers with different preference may have different arrival time. For instance, in the airline revenue management problems, casual travelers, whose reserve prices are probably lower, usually arrive long before their scheduled departure time; while business travelers, who tend to be price-insensitive, are more likely to appear shortly before intended trips. In this section, we take this heterogeneity into account.

Assume all customers are categorized into $K$ groups: $N = \bigcap_{k=1}^{K} N_k$. Furthermore, we assume there exists a constant $c$ such that $\forall k, k', t, F_k(t) \leq cF_{k'}(t)$. Let $t_0$ be the $\epsilon$-quantile point, i.e. $F_1(t_0) = \epsilon$. Assume $F_k(t_0) = r_k \epsilon$. From the assumption on the CDFs, we have $r_k \in [1/c, c]$.

Let $S_k$ be the set of customers from group $k$ that arrive before $t_0$. The online algorithm observes customers arriving before $t_0$ and solves the following LPs:

$$
\begin{array}{llll}
\max & \sum_k (r_k \epsilon)^{-1} \sum_{j \in S_k}^{o \in O_j} \pi_{jo}^k x_{jo}^k & \min & \epsilon(1-\epsilon) \sum_i b_i p_i + \sum_{j,k} q_j^k \\
s.t. & \sum_k (r_k \epsilon)^{-1} \sum_{j \in S_k}^{o \in O_j} a_{ijo}^k x_{jo}^k \leq \epsilon(1-\epsilon)b_i, \quad \forall i & s.t. & r_k \epsilon q_j^k + \sum_i a_{ijo}^k p_i \geq \pi_j^k, \forall j, o, k \\
& \sum_{o \in O_j} x_{jo}^k \leq 1, \forall j, k & & \mathbf{p}, \mathbf{q} \geq 0 \\
& \mathbf{x} \geq 0
\end{array}
\tag{11}
$$

Similar to arguments in the previous sections, We show that with high probability, the resulting solution is feasible, the resulting solution is near optimal, and the loss due to observation is small:

**Lemma 12.** *If $\min_i b_i \geq 3cm \ln(nq/\epsilon)/\epsilon^3$, then w.p. $1-\epsilon$, $\forall i, \sum_k \sum_{j \in N_k}^{o \in O_j} a_{ijo}^k x_{jo}^k(\hat{\mathbf{p}}) \leq b_i$.*

**Lemma 13.** *If $\min_i b_i \geq 3cm \ln(nq/\epsilon)/\epsilon^3$, then w.p. $1-\epsilon$, $\sum_k \sum_{j \in N_k}^{o \in O_j} \pi_{jo}^k x_{jo}^k(\hat{\mathbf{p}}) \geq (1-3\epsilon)OPT$.*

**Lemma 14.** $\mathbb{E}[\sum_k \sum_{j \in S_k}^{o \in O_j} \pi_{jo}^k x_{jo}^k] \leq \max_k r_k \epsilon OPT \leq c\epsilon OPT$.

Combining the three lemmas above, we can conclude that:

**Theorem 4.** *If $\min_i b_i \geq 3cm \ln(nq/\epsilon)/\epsilon^3$, the algorithm is $1 - O(\epsilon)$ competitive.*

Unfortunately, dynamic pricing techniques used in DPA and DPAD do not apply for this problem, because the arrival process is not homogeneous here. Without dynamic pricing mechanism, in order to have near optimality result, the lower bound imposed on the model is much higher than the one we obtained in the previous sections. Worth noting that, this approach can only deal with problems where customers of each type are well represented in the early stages. Otherwise, we may need additional assumptions of information to obtain good results. For example, consider airline tickets sales, if all casual travelers arrive at least one week before departure and all business travelers only appear one week within departure. If the numbers of travelers of the two types are unrelated, then past information alone is unlikely to help us decide how many seats to reserve for business customers. To find a proper reserve level, we need good estimates on the numbers of travelers of the two types, which probably requires more assumptions.

# A    Proof of Lemma 1

**Lemma 1.** *For non-degenerate $j \in S_\epsilon$, $\hat{x}_{jo} = x_{jo}(\hat{\mathbf{p}})$ for all $o \in O_j$.*

*Proof.* If there exists $o \in O_j$ such that $\hat{x}_{jo} > 0$. From complementary slackness, we have $\sum_i a_{ijo}\hat{p}_i + \hat{q}_j = \pi_{jo}$. Since $\forall o' \in O_j$,

$$\pi_{jo'} - \sum_i a_{ijo'}\hat{p}_i < \hat{q}_j = \pi_{jo} - \sum_i a_{ijo}\hat{p}_i,$$

we have $x_{jo}(\hat{\mathbf{p}}) = 1$ and $\hat{q}_j > 0$. Since $j$ is non-degenerate, then $\hat{q}_j > 0$. Combined with complementary slackness, we have $\sum_{o' \in O_j} \hat{x}_{jo'} = 1$. Note that $\forall o' \neq o$, $\hat{x}_{jo'} = 0$ because $\pi_{jo'} - \sum_i a_{ijo'}\hat{p}_i < \hat{q}_j$. Hence, $\hat{x}_{jo} = 1 = x_{jo}(\hat{\mathbf{p}})$.

If $\hat{x}_{jo} = 0$ for all $o \in O_j$, then $\hat{q}_j = 0$. Since $\pi_{jo} - \sum_i a_{ijo}\hat{p}_i \leq \hat{q}_j$ and $j$ is non-degenerate, $\pi_{jo} - \sum_i a_{ijo}\hat{p}_i < 0$. Therefore, $x_{jo}(\hat{\mathbf{p}}) = 0$ for all $o \in O_j$. $\qquad\blacksquare$

# B    Omitted Proofs in Section 3

**Lemma 5.** *If $\min_i b_i \geq 10m\ln(nq/\epsilon)/\epsilon^2$, then w.p. $1 - \epsilon$, $\sum_{j \in S_{2l} \setminus S_l, o \in O_j} a_{ijo}x_{jo}(\hat{\mathbf{p}}_l) \leq lb_i, \forall i, l$.*

*Proof.* The proof is very similar to the one of Lemma 2. Let us first consider the probability of event

$$\{\sum_{j \in S_l}^{o \in O_j} a_{ijo}x_{jo}(\mathbf{p}_l) \leq (1 - h_l)lb_i, \sum_{j \in S_{2l} \setminus S_l}^{o \in O_j} a_{ijo}x_{jo}(\mathbf{p}_l) \geq lb_i\} \tag{12}$$

for all fixed $l$, $\mathbf{p}_l$, and $i$:

$$\begin{aligned}
&\Pr(\sum_{j \in S_l, o \in O_j} a_{ijo}x_{jo}(\mathbf{p}_l) \leq (1-h_l)lb_i, \sum_{j \in S_{2l} \setminus S_l, o \in O_j} a_{ijo}x_{jo}(\mathbf{p}_l) \geq lb_i)\\
\leq\ &\Pr(\sum_{j \in S_l, o \in O_j} a_{ijo}x_{jo}(\mathbf{p}_l) \leq (1-h_l)lb_i, \sum_{j \in S_{2l}, o \in O_j} a_{ijo}x_{jo}(\mathbf{p}_l) \geq (2-h_l)lb_i)\\
+\ &\Pr(\sum_{j \in S_{2l}, o \in O_j} a_{ijo}x_{jo}(\mathbf{p}_l) \geq lb_i, \sum_{j \in S_{2l}, o \in O_j} a_{ijo}x_{jo}(\mathbf{p}_l) \leq (2-h_l)lb_i).
\end{aligned}$$

Note that $\Pr(j \in S_l | j \in S_{2l}) = 1/2$. From Bernstein inequalities, the first term is upper bounded by $\exp(-\epsilon^2 b_i/10)$. Similarly, the second term is also upper bounded by $\exp(-\epsilon^2 b_i/10)$. Thus,

$$\Pr(\sum_{j \in S_l}^{o \in O_j} a_{ijo}x_{jo}(\mathbf{p}_l) \leq (1-h_l)lb_i, \sum_{j \in S_{2l} \setminus S_l}^{o \in O_j} a_{ijo}x_{jo}(\mathbf{p}_l) \geq lb_i) \leq 2\exp(-\epsilon^2 b_i/10).$$

Note that for each $l$, there are at most $(nq)^m$ distinct $\mathbf{p}_l$ regions. By union bounds, we have that with probability $\epsilon$, there exist $i$, $l$, and $\mathbf{p}_l$, such that (12) is true. On the other hand, from Lemma 1, we have $\sum_{j \in S_l, o \in O_j} a_{ijo}x_{jo}(\hat{\mathbf{p}}) \leq \sum_{j \in S_l}^{o \in O_j} a_{ijo}\hat{x}_{jo} \leq (1-h_l)lb_i$. By letting $\mathbf{p}_l = \hat{\mathbf{p}}_l$, we can conclude our lemma. $\qquad\blacksquare$

**Lemma 6.** *If $\min_i b_i \geq 10m\ln(nq/\epsilon)/\epsilon^2$, then w.p. $1 - \epsilon$, $\sum_{j \in S_{2l}, o \in O_j} \pi_{jo}x_{jo}(\hat{\mathbf{p}}_l) \geq (1 - 2h_l - \epsilon)OPT_{2l}, \forall l$.*

*Proof.* Let us consider the following LP:

$$\begin{aligned}
\max\quad & \sum_{j \in S_{2l}, o} \pi_{jo}x_{jo}\\
s.t.\quad & \sum_{j \in S_{2l}, o} a_{ijo}x_{jo} \leq \hat{b}_i, && \forall i\\
& \sum_{o \in O_j} x_{jo} \leq 1, && \forall j \in S_{2l}\\
& x_{jo} \geq 0, && \forall j \in S_{2l}, o
\end{aligned} \tag{13}$$

where $\hat{b}_i = \sum_{j \in S_{2l}} a_{ijo} x_{jo}(\hat{\mathbf{p}}_\mathbf{l})$, if $\hat{p}_{l,i} > 0$ and $\hat{b}_i = \max\{\sum_{j \in 2l} a_{ijo} x_{jo}(\hat{\mathbf{p}}_\mathbf{l}), b_i\}$, if $\hat{p}_{l,i} = 0$. By complementary slackness, $\mathbf{x}(\hat{\mathbf{p}}_\mathbf{l})$ is the optimal solution to this LP.

On the other hand, using the same argument as in the proof of Lemma 3, we can show that with probability $1 - \epsilon$, $\hat{b}_i \geq 2l \cdot (1 - 2h_l - \epsilon) b_i$ for all $i$ and $l$. In that case, $\sum_{j \in S_{2l,o}} \pi_{jo} x_{jo}(\hat{\mathbf{p}}_\mathbf{l}) \geq (1 - 2h_l - \epsilon) OPT_{2l}$. $\qquad\square$

**Lemma 7.** $\mathbb{E}[OPT_l] \leq l \cdot OPT$.

*Proof.* Consider the optimal dual solution $(\mathbf{p}^*, \mathbf{q}^*)$ to LP (1). We can easily check that it is feasible to the dual problem of LP (9):

$$
\begin{aligned}
\min \quad & \sum_i (1 - h_l) l \epsilon b_i p_i + \sum_{j \in S_l} q_j \\
s.t. \quad & \sum_i a_{ijo} p_i + q_j \geq \pi_{jo}, \forall j \in S_l, o \in O_j \\
& p_i \geq 0, \forall i \\
& q_j \geq 0, \forall j \in S_l
\end{aligned}.
$$

Thus, $OPT_l \leq (1 - h_l) l \sum_i b_i p_i^* + \sum_{j \in S_l} q_j^*$. Note that for every $j$, $\Pr(j \in S_l) = l$. By taking expectation on both sides, we can conclude the lemma. $\qquad\square$

**Theorem 2.** *If* $\min_i b_i \geq 10m \ln(nq/\epsilon)/\epsilon^2$, *then DPA is* $1 - O(\epsilon)$ *competitive.*

*Proof.* Let $\mathcal{E}_2$ denote the event that both inequalities in Lemma 5 and 6 are true, then $\Pr(\mathcal{E}_2) \geq 1 - 2\epsilon$.

$$
\begin{aligned}
& \mathbb{E}[\sum_{l \in L} \sum_{j \in S_{2l} \setminus S_l} \pi_j x_j(\hat{\mathbf{p}}_\mathbf{l}) | \mathcal{E}_2] \\
\geq \quad & \sum_{l \in L} \mathbb{E}[\sum_{j \in S_{2l}} \pi_j x_j(\hat{\mathbf{p}}_\mathbf{l}) | \mathcal{E}_2] - \sum_{l \in L} \mathbb{E}[\sum_{j \in S_l} \pi_j x_j(\hat{\mathbf{p}}_\mathbf{l}) | \mathcal{E}_2] \\
\geq \quad & \sum_{l \in L} (1 - 2h_l - \epsilon) \mathbb{E}[OPT_{2l} | \mathcal{E}_2] - \sum_{l \in L} \mathbb{E}[OPT_l | \mathcal{E}_2] \\
\geq \quad & OPT - \sum_{l \in L} 2h_l \mathbb{E}[OPT_{2l} | \mathcal{E}_2] - \epsilon \sum_{l \in L} \mathbb{E}[OPT_{2l} | \mathcal{E}_2] - \mathbb{E}[OPT_\epsilon | \mathcal{E}_2] \\
\geq \quad & OPT - 4 \sum_{l \in L} h_l l \cdot OPT - 2\epsilon \sum_{l \in L} l \cdot OPT - \epsilon OPT \\
\geq \quad & OPT - 13\epsilon OPT
\end{aligned}.
$$

Therefore, $\mathbb{E}[\sum_{l \in L} \sum_{j \in S_{2l} \setminus S_l} \pi_j x_j(\hat{\mathbf{p}}_\mathbf{l})] \geq (1 - 15\epsilon) OPT$. $\qquad\square$

# C  Omitted Proofs in Section 4

**Lemma 9.** *Given* $\mathcal{E}_{est}$, *if* $\min_i b_i \geq 3m \ln(nq/\epsilon)/\epsilon^2$, *then with probability* $1 - \epsilon$,

$$
\sum_{j \in S_{2l} \setminus S_l, o \in O_j} a_{ijo} x_{jo}(\hat{\mathbf{p}}_\mathbf{l}) \leq lb_i, \forall i, l.
$$

*Proof.* Fix $\hat{\mathbf{p}}$, $i$, and $l$. Let $X_j = \sum_{o \in O_j} a_{ijo} x_{jo}(\hat{\mathbf{p}})$. Then,

$$
\begin{aligned}
& \Pr(\sum_{j \in S_{2l} \setminus S_l} X_j > lb_i, \sum_{j \in S_l} X_j \leq (1 - 6h_l) lb_i) \\
\leq \quad & \Pr(\sum_{j \in S_l} X_j \leq (1 - 6h_l) lb_i, \sum_{j \in S_{2l}} X_j \geq 2(1 - 3h_l) lb_i) \\
& + \Pr(\sum_{j \in S_{2l} \setminus S_l} X_j \geq lb_i, \sum_{j \in S_{2l}} X_j \leq 2(1 - 3h_l) lb_i)
\end{aligned}.
$$

Since $\Pr(j \in S_l | j \in S_{2l}) = F(t'_{lk})/F(t'_{2lk}) \leq (1 + 2h_l)/2$, the first term

$$\Pr(\sum_{j \in S_l} X_j \leq (1 - 6h_l)lb_i, \sum_{j \in S_{2l}} X_j \geq 2(1 - 3h_l)lb_i)$$
$$\leq \Pr(\sum_{j \in S_l} X_j - \mathbb{E}[\sum_{j \in S_l} X_j] \leq \min\{-h_l \sum_{j \in S_{2l}} X_j/2, -h_l lb_i\}, \sum_{j \in S_{2l}} X_j \geq 2(1 - 3h_l)lb_i) .$$
$$\leq \exp(-\epsilon^2 b_i/3)$$

Since $\Pr(j \in S_{2l}\backslash S_l | j \in S_{2l}) = 1 - \Pr(j \in S_l | j \in S_{2l}) \geq (1 - 2h_l)/2$, the second term

$$\Pr(\sum_{j \in S_{2l}\backslash S_l} X_j \geq lb_i, \sum_{j \in S_{2l}} X_j \leq 2(1 - 3h_l)lb_i)$$
$$\leq \Pr(\sum_{j \in S_{2l}\backslash S_l} X_j - \mathbb{E}[\sum_{j \in S_{2l}\backslash S_l} X_j] \geq h_l lb_i, \sum_{j \in S_{2l}} X_j \leq 2(1 - 3h_l)lb_i) .$$
$$\leq \exp(-\epsilon^2 b_i/3)$$

Therefore, $\Pr(\sum_{j \in S_{2l}\backslash S_l} X_j > lb_i, \sum_{j \in S_l} X_j \leq (1 - 6h_l)lb_i) \leq 2\exp(-\epsilon^2 b_i/3)$. By taking union bounds over all possible $\mathbf{p_l}$, $i$, and $l$, we can conclude the lemma. □

**Lemma 10.** *Given $\mathcal{E}_{est}$, if $\min_i b_i \geq 3m \ln(nq/\epsilon)/\epsilon^2$, then with probability $1 - \epsilon$,*

$$\sum_{j \in S_{2l}, o \in O_j} \pi_{jo} x_{jo}(\hat{\mathbf{p}}_\mathbf{l}) \geq (1 - 9h_l - \epsilon)OPT_{2l}, \forall l.$$

*Proof.* Let us consider the following LP:

$$\begin{array}{ll}
\max & \sum_{j \in S_{2l}, o} \pi_{jo} x_{jo} \\
s.t. & \sum_{j \in S_{2l}, o} a_{ijo} x_{jo} \leq \hat{b}_i, \quad \forall i \\
& \sum_{o \in O_j} x_{jo} \leq 1, \quad \forall j \in S_{2l} \\
& x_{jo} \geq 0, \quad \forall j \in S_{2l}, o
\end{array} \quad (14)$$

where $\hat{b}_i = \sum_{j \in S_{2l}} a_{ijo} x_{jo}(\hat{\mathbf{p}}_\mathbf{l})$, if $\hat{p}_{l,i} > 0$ and $\hat{b}_i = \max\{\sum_{j \in S_{2l}} a_{ijo} x_{jo}(\hat{\mathbf{p}}_\mathbf{l}), b_i\}$, if $\hat{p}_{l,i} = 0$. By complementary slackness, $\mathbf{x}(\hat{\mathbf{p}}_\mathbf{l})$ is the optimal solution to this LP.

On the other hand, using the same argument as in the proof of Lemma 3, we can show that with probability $1 - \epsilon$, $\hat{b}_i \geq 2l \cdot (1 - 9h_l - \epsilon)b_i$ for all $i$ and $l$. In that case, then $\sum_{j \in S_{2l}, o} \pi_{jo} x_{jo}(\hat{\mathbf{p}}_\mathbf{l}) \geq (1 - 9h_l - \epsilon)OPT_{2l}$. □

**Lemma 11.** *Given $\mathcal{E}_{est}$, $\mathbb{E}[OPT_l] \leq (1 + h_l)l \cdot OPT$.*

*Proof.* Consider the optimal dual solution $(\mathbf{p}^*, \mathbf{q}^*)$ to LP (1). We can easily check that it is feasible to the partial dual problem (9)

$$\begin{array}{ll}
\min & \sum_i (1 - 6h_l)l\epsilon b_i p_i + \sum_{j \in S_l} q_j \\
s.t. & \sum_i a_{ijo} p_i + q_j \geq \pi_{jo}, \quad \forall j \in S_l, o \in O_j \\
& p_i \geq 0, \forall i \\
& q_j \geq 0, \forall j \in S_l
\end{array} .$$

Thus, $OPT_l \leq (1 - 6h_l)l \sum_i b_i p_i^* + \sum_{j \in S_l} q_j^*$. Note that given $\mathcal{E}_{est}$, for every $j$, $\Pr(j \in S_l) \leq (1 + h_l)l$. By taking expectation on both sides, we can conclude the lemma. □

**Theorem 3.** *If $\min_i b_i \geq 3m \ln(nq/\epsilon)/\epsilon^2$ and $k \geq 5 \ln \epsilon/\epsilon^2$, the algorithm is $1 - O(\epsilon)$ competitive.*

*Proof.* Let $\mathcal{E}_3$ denote the event that both inequalities in Lemma 9 and 10 are true, then $\Pr(\mathcal{E}_3 \cap \mathcal{E}_{est}) \geq 1 - 3\epsilon$.

$$
\begin{aligned}
& \mathbb{E}[\sum_{l \in L} \sum_{j \in S_{2l} \setminus S_l} \pi_j x_j(\hat{\mathbf{p}}_1) | \mathcal{E}_3 \cap \mathcal{E}_{est}] \\
\geq & \sum_{l \in L} \mathbb{E}[\sum_{j \in S_{2l}} \pi_j x_j(\hat{\mathbf{p}}_1) | \mathcal{E}_3 \cap \mathcal{E}_{est}] - \sum_{l \in L} \mathbb{E}[\sum_{j \in S_l} \pi_j x_j(\hat{\mathbf{p}}_1) | \mathcal{E}_3 \cap \mathcal{E}_{est}] \\
\geq & \sum_{l \in L} (1 - 9h_l - \epsilon) \mathbb{E}[OPT_{2l} | \mathcal{E}_3 \cap \mathcal{E}_{est}] - \sum_{l \in L} \mathbb{E}[OPT_l | \mathcal{E}_3 \cap \mathcal{E}_{est}] \\
\geq & \ OPT - \sum_{l \in L} 9h_l \mathbb{E}[OPT_{2l} | \mathcal{E}_3 \cap \mathcal{E}_{est}] - \epsilon \sum_{l \in L} \mathbb{E}[OPT_{2l} | \mathcal{E}_3 \cap \mathcal{E}_{est}] - \mathbb{E}[OPT_\epsilon | \mathcal{E}_3 \cap \mathcal{E}_{est}] \\
\geq & \ OPT - 9 \sum_{l \in L} (2h_l l + h_l h_{2l} l) \cdot OPT - \epsilon \sum_{l \in L} 4l \cdot OPT - 2\epsilon OPT \\
\geq & \ OPT - 42\epsilon OPT
\end{aligned}
$$

Therefore, $\mathbb{E}[\sum_{l \in L} \sum_{j \in S_{2l} \setminus S_l} \pi_j x_j(\hat{\mathbf{p}}_1)] \geq (1 - 45\epsilon) OPT$. $\qquad\blacksquare$

# D  Omitted Proofs in Section 5

**Lemma 12.** *If* $\min_i b_i \geq 3cm \ln(nq/\epsilon)/\epsilon^3$, *then w.p.* $1 - \epsilon, \forall i, \sum_k \sum_{j \in N_k}^{o \in O_j} a_{ijo}^k x_{jo}^k(\hat{\mathbf{p}}) \leq b_i$.

*Proof.* The proof is very similar to the one of Lemma 2. Let us first consider the probability of the event

$$
\{\sum_k (r_k \epsilon)^{-1} \sum_{j \in S_k}^{o \in O_j} a_{ijo}^k x_{jo}^k(\mathbf{p}) \leq \epsilon(1 - \epsilon) b_i, \sum_k \sum_{j \in N_k}^{o \in O_j} a_{ijo}^k x_{jo}^k(\mathbf{p}) \geq b_i\} \tag{15}
$$

for all fixed $\mathbf{p}$ and $i$. Since $\Pr(j \in S_k | j \in N_k) = r_k \epsilon$, we expect $\sum_k (r_k \epsilon)^{-1} \sum_{j \in S_k}^{o \in O_j} a_{ijo}^k x_{jo}^k$ close to its mean $\sum_k \sum_{j \in N_k}^{o \in O_j} a_{ijo}^k x_{jo}^k$. Therefore, event (15) should be a rare event. More precisely,

$$
\Pr(\sum_k (r_k \epsilon)^{-1} \sum_{j \in S_k}^{o \in O_j} a_{ijo}^k x_{jo}^k(\mathbf{p}) \leq \epsilon(1 - \epsilon) b_i, \sum_k \sum_{j \in N_k}^{o \in O_j} a_{ijo}^k x_{jo}^k(\mathbf{p}) \geq b_i) \leq \exp(-\epsilon^3 b_i/3c).
$$

Note that there are at most $(nq)^m$ distinct $\mathbf{p}$. By taking union bounds over all distinct $\mathbf{p}$ and $i$, we can conclude the lemma. $\qquad\blacksquare$

**Lemma 13.** *If* $\min_i b_i \geq 3cm \ln(nq/\epsilon)/\epsilon^3$, *then w.p.* $1 - \epsilon, \sum_k \sum_{j \in N_k}^{o \in O_j} \pi_{jo}^k x_{jo}^k(\hat{\mathbf{p}}) \geq (1 - 3\epsilon) OPT$.

*Proof.* Let us consider the following LP:

$$
\begin{aligned}
\max \quad & \sum_k \sum_{j \in N_k}^{o \in O_j} \pi_{jo}^k x_{jo}^k \\
s.t. \quad & \sum_k \sum_{j \in N_k}^{o \in O_j} a_{ijo}^k x_{jo}^k \leq \hat{b}_i, \quad \forall i \\
& \sum_{o \in O_j} x_{jo}^k \leq 1, \qquad\qquad \forall j, k \\
& x_{jo}^k \geq 0, \qquad\qquad\qquad \forall j, k, o
\end{aligned} \tag{16}
$$

where $\hat{b}_i = \sum_k \sum_{j \in N_k}^{o \in O_j} a_{ijo}^k x_{jo}^k(\hat{\mathbf{p}})$, if $\hat{p}_i > 0$ and $\hat{b}_i = \max\{\sum_k \sum_{j \in N_k}^{o \in O_j} a_{ijo}^k x_{jo}^k(\hat{\mathbf{p}}), b_i\}$, if $\hat{p}_i = 0$. By complementary slackness, $\mathbf{x}(\hat{\mathbf{p}})$ is the optimal solution to this LP.

On the other hand, using the same argument as in the proof of Lemma 2, we can show that with probability $1 - \epsilon$, $\hat{b}_i \geq (1 - 3\epsilon) b_i$ for all $i$ and $l$. If such an event happens, then $\sum_k \sum_{j \in N_k}^{o \in O_j} \pi_{jo}^k x_{jo}^k(\hat{\mathbf{p}}) \geq (1 - 3\epsilon) OPT$. $\qquad\blacksquare$

**Lemma 14.** $\mathbb{E}[\sum_k \sum_{j\in S_k}^{o\in O_j} \pi_{jo}^k x_{jo}^k] \le \max_k r_k \epsilon OPT \le c\epsilon OPT$.

*Proof.* Let $OPT_\epsilon$ be the optimal value to the partial LP (11). Let $(\mathbf{p}^*, \mathbf{q}^*)$ be the optimal dual solution to the complete LP (16). It is easy to check that $(\mathbf{p}^*, \mathbf{q}^*)$ is a dual feasible solution to (11). Therefore, $\mathbb{E}[OPT_\epsilon] \le OPT$.

On the other hand, for any realization, the lost revenue resulting from the first $\epsilon$ fraction customers is no more than $OPT_1$. Hence,

$$\mathbb{E}[\sum_k \sum_{j\in S_k}^{o\in O_j} \pi_{jo}^k x_{jo}^k] \le \mathbb{E}[\max_k r_k \epsilon \cdot OPT_\epsilon] \le \max_k r_k \epsilon \cdot OPT.$$

$\square$

**Theorem 4.** *If $\min_i b_i \ge 3cm \ln(nq/\epsilon)/\epsilon^3$, the algorithm is $1 - O(\epsilon)$ competitive.*

*Proof.* Let $\mathcal{E}_4$ denote the event that for all $i$

$$\sum_k \sum_{j\in N_k}^{o\in O_j} a_{ijo}^k x_{jo}^k(\hat{\mathbf{p}}) \le b_i$$

and

$$\sum_k \sum_{j\in N_k}^{o\in O_j} \pi_{jo}^k x_{jo}^k(\hat{\mathbf{p}}) \ge (1 - 3\epsilon)OPT.$$

Then, from Lemma 12 and 13, we have $\Pr(\mathcal{E}_4) \le 1 - 2\epsilon$. Given $\mathcal{E}$, the online solution $\mathbf{x}(\hat{\mathbf{p}})$ is feasible. Therefore,

$$
\begin{aligned}
\mathbb{E}[\sum_k \sum_{j\notin S_k}^{o\in O_j} a_{ijo}^k x_{jo}^k(\hat{\mathbf{p}})] &\ge \mathbb{E}[\sum_k \sum_{j\notin S_k}^{o\in O_j} a_{ijo}^k x_{jo}^k(\hat{\mathbf{p}})|\mathcal{E}_4] \cdot \Pr(\mathcal{E}_4) \\
&\ge (\mathbb{E}[\sum_k \sum_{j\in N_k}^{o\in O_j} a_{ijo}^k x_{jo}^k(\hat{\mathbf{p}})|\mathcal{E}_4] - \mathbb{E}[\sum_k \sum_{j\in S_k}^{o\in O_j} \pi_{jo}^k x_{jo}^k|\mathcal{E}_4]) \cdot \Pr(\mathcal{E}_4) \\
&\ge (1 - 3\epsilon)(1 - 2\epsilon)OPT - c\epsilon OPT \\
&\ge (1 - O(\epsilon))OPT
\end{aligned}
$$

$\square$

# References

[1] G. Aggarwal, G. Goel, C. Karande, and A. Mehta. Online vertex-weighted bipartite matching and single-bid budgeted allocations. In *SODA '11: Proceedings of the 22nd Annual ACM-SIAM Symposium on Discrete Algorithms*, pages 1253–1264, 2011.

[2] S. Agrawal, Z. Wang, and Y. Ye. A dynamic near-optimal algorithm for online linear programming. Working paper, Stanford University, 2009.

[3] M. Babaioff, N. Immorlica, D. Kempe, and R. Kleinberg. Online auctions and generalized secretary problems. *SIGecom Exchanges*, 7(2):1–11, June 2008.

[4] S. Bernstein. On a modification of Chebyshev's inequality and of the error formula of Laplace. *Ann. Sci. Inst. Sav. Ukraine, Sect. Math.*, 1:38–49, 1924.

[5] N. Buchbinder, K. Jain, and J. Noar. Online primal-dual algorithms for maximizing ad-auctions revenue. In *ESA'07 Proceedings of the 15th Annual European Conference on Algorithms*, pages 253–264, 2007.

[6] N. Devanur and T. Hayes. The adwords problem: online keyword matching with budgeted bidders under random permutation. In *EC'09: Proceedings of the 10th ACM conference on Electronic Commerce*, pages 71–78, 2009.

[7] J. Feldman, M. Henzinger, N. Korula, V. Mirrokni, and C. Stein. Online stochastic packing applied to display ad allocation. In *ESA'10 Proceedings of the 18th Annual European Conference on Algorithms*, pages 182–194, 2010.

[8] J. Feldman, A. Mehta, V. Mirrokni, and S. Muthukrishnan. Online stochastic matching: beating 1-1/e. In *FOCS'09: Proceedings of the 50th Annual IEEE Symposium on Foundations of Computer Science*, pages 117–126, 2009.

[9] G. Goel and A. Mehta. Online budgeted matching in random input models with applications to adwords. In *SODA'08: Proceedings of the 16th Annual ACM-SIAM Symposium on Discrete Algorithms*, pages 982–991, 2008.

[10] P. Jaillet and X. Lu. Online stochastic matching: New algorithms with better bounds. Working paper, MIT, 2012.

[11] C. Karande, A. Mehta, and P. Tripathi. Online bipartite matching with unknown distributions. In *STOC'11 Proceedings of the 43rd Annual ACM Symposium on Theory of Computing*, pages 587–596, 2011.

[12] R. Karp, U. Vazirani, and V. Vazirani. An optimal algorithm for online bipartite matching. In *STOC'90 Proceedings of the 22nd Annual ACM Symposium on Theory of Computing*, pages 352–358, 1990.

[13] M. Mahdian and Q. Yan. Online bipartite matching with random arrivals: a strongly factor revealing lp approach. In *STOC'11 Proceedings of the 43rd Annual ACM Symposium on Theory of Computing*, pages 597–606, 2011.

[14] V. Manshadi, S. Oveis Gharan, and A. Saberi. Online stochastic matching: online actions based on offline statistics. In *SODA'11 Proceedings of the 22nd Annual ACM-SIAM Symposium on Discrete Algorithms*, pages 1285–1294, 2011.

[15] A. Mehta, A. Saberi, U. Vazirani, and V. Vazirani. Adwords and generalized online matching. In *FOCS'05: Proceedings of the 4th Annual IEEE Symposium on Foundations of Computer Science*, pages 264–273, 2005.

[16] V. Mirrokni, S. Oveis Gharan, and M. Zadimoghaddam. Simultaneous approximations for adversarial and stochastic online budgeted allocation. In *SODA'12 Proceedings of the 23rd Annual ACM-SIAM Symposium on Discrete Algorithms*, pages 1690–1701, 2012.

[17] P. Orlik and H. Terao. *Arrangement of Hyperplanes*. Springer-Verlag, 1992.